\documentclass[a4paper]{jpconf}
\usepackage{graphicx}
\usepackage{amssymb}
\usepackage{amsmath}

\newcommand{\be}{\begin{equation}} 
\newcommand{\ee}{\end{equation}} 
\newcommand{\ba}{\begin{eqnarray}} 
\newcommand{\ea}{\end{eqnarray}} 

\begin{document}
\title{Strange and heavy mesons in hadronic matter}

\author{Daniel Cabrera$^{1,}\footnote[0]{Speaker}$, Luciano M Abreu$^2$, Elena Bratkovskaya$^1$, Andrej Ilner$^1$, Felipe J Llanes-Estrada$^3$, Angels Ramos$^4$, Laura Tolos$^{1,5}$ and Juan M Torres-Rincon$^5$}

\address{
$^1$ Institute for Theoretical Physics and Frankfurt Institute for Advanced Studies, Frankfurt University, 60438 Frankfurt am Main, Germany\\
$^2$ Instituto de F\'{\i}sica, Universidade Federal da Bahia, 40210-340. Salvador, BA, Brazil \\
$^3$ Departamento de F\'{\i}sica Te\'orica I, Universidad Complutense, 28040 Madrid, Spain\\
$^4$ Departament d'Estructura i Constituents de la Mat\`eria, Universitat de Barcelona, Diagonal 647, 08028 Barcelona, Spain\\
$^5$ Institut de Ci\`encies de l'Espai (IEEC/CSIC), Campus Universitat Aut\`onoma de Barcelona, Facultat de Ci\`encies, Torre C5, E-08193 Bellaterra, Spain
}

%\ead{cabrera@fias.uni-frankfurt.de}

\begin{abstract}
We present selected results on the properties of strange and heavy-flavoured mesons in a hot and dense nuclear medium, with emphasis in selfconsistent coupled-channel approaches based on the chiral Lagrangian. In the strangeness sector, we discuss how the enhanced reactivity of light strange vectors at FAIR conditions can be tied to in-medium effects on their predominant decay modes (e.g.~$\bar K^*\to\bar K \pi$) and to the excitation of strange baryons in vector-meson nucleon interactions. In the heavy-flavour sector, we focus on recent determinations of the transport coefficients of charmed and bottomed mesons in a hadron gas at vanishing baryonic chemical potential. We comment on the role of microscopic transport simulations to establish a connection between theoretical models and experimental observables from heavy-ion collisions (HICs).
\end{abstract}

\section{Introduction}
The features of strongly interacting matter in a broad range of temperature and density has been a subject of great interest in the last decades, in connection with fundamental aspects of the strong interaction such as the nature of (de)confinement or the physical mechanism of chiral symmetry restoration, and with the microscopic dynamics and composition of neutron stars.

In particular, understanding the dynamics of light strange mesons in a nuclear environment has posed a challenge to theoretical models, amongst different reasons, due to the failure of the low-density approximation to describe the interaction of the $\bar K$ meson with the medium, as it is concluded from the phenomenology of kaonic atoms \cite{Friedman:2007zza}.
The properties of both $K$ and $\bar K$ close to threshold energy have also been thoroughly investigated in HICs at SIS energies \cite{Fuchs:2005zg,Forster:2007qk,Hartnack:2011cn,Cassing:2003vz}. The analysis of experimental data in conjunction with microscopic transport approaches has allowed to draw solid conclusions regarding the production mechanisms of strangeness, the freeze-out conditions exhibited by $K^+$ and $K^-$ mesons and the use of $K^+$ as a probe of the nuclear matter equation of state at high densities. A good agreement with data has been achieved for many different observables when the in-medium properties of kaons are implemented. Still, a simultaneous description of all observables involving antikaon production is still missing, leaving room for a more elaborated description of the most relevant reactions (e.g., $\pi Y \to \bar K N$).

In the heavy flavour sector, the study of the mechanisms of equilibration of heavy mesons in a hadronic medium has been lately subject of intense activity. This has been partially motivated by recent developments on the theoretical description of the interaction of $D$ and $B$ mesons with other (lighter) hadrons, based on phenomenological approaches \cite{He:2011yi} or effective theories of QCD \cite{Ghosh:2011bw,Abreu:2011ic,Das:2011vba,Abreu:2012et}. The effect of these interactions in the heavy meson spectrum is quantified by the corresponding transport coefficients. 

We report in this talk on recent progress on the medium modifications of strange and heavy flavoured mesons, highlighting (some new) results from chirally motivated hadronic many-body approaches in connection with microscopic transport simulations.

\section{Strange mesons in a hot/dense medium}

\subsection{Selfconsistent determination of $\bar K$ and $K$ spectral functions at FAIR conditions}

The early departure of the $\bar K$ nuclear potential from the low-density approximation is successfully described in coupled-channel approaches implementing the exact unitarity of the scattering matrix and a selfconsistent evaluation of the kaon selfenergy \cite{Ramos:1999ku,Tolos:2000fj}. The onset of an attractive $s$-wave $\bar K N$ interaction at low densities is a consequence of an upper shift of the $\Lambda(1405)$ resonance induced by Pauli blocking on the nucleon states. The simultaneous dressing of all intermediate meson-baryon states in such approaches typically leads to an attractive $\bar K$ potential of about 40-60~MeV at normal nuclear matter density ($\rho_0$), a rather shallower potential than obtained within relativistic mean-field calculations \cite{Schaffner:1996kv} or required by phenomenological analysis of kaonic atom data \cite{Friedman:2007zza}.

In Ref.~\cite{Tolos:2008di} the coupled-channel approach of \cite{Ramos:1999ku}, based on the meson-baryon chiral Lagrangian, was extended to account for finite temperature so as to match the experimental conditions at the future FAIR facility at GSI. 
The set of coupled equations determining the $\bar K (K) N$ scattering amplitudes and the $\bar K (K)$ selfenergies were solved selfconsistently for the lowest partial waves (schematically $\Pi_{\bar K(K)}=\sum_{\vec{p}}n(\vec{p}\,) T_{\bar K(K)N}$) accounting for Pauli-blocking effects, finite-temperature baryonic mean-field potentials and the meson selfenergies of pions and kaons.
The model achieves a good description of vacuum low-energy scattering observables with a minimal number of parameters. The $\Lambda (1405)$ resonance is dynamically generated in the isospin I=0 channel from the leading order (Weinberg-Tomozawa) interaction, once unitarity is restored by solving the Bethe-Salpeter equation for the scattering amplitudes.
A $p$-wave selfenergy for the kaons was implemented accounting for $YN^{-1}$ excitations, where $Y=\Lambda, \Sigma, \Sigma^*$ (see also \cite{Lutz:2007bh}). The role of higher partial waves (beyond $L=0$), leading to non-isotropic in-medium cross sections, has been emphasised in the context of heavy-ion collisions \cite{Fuchs:2005zg}.

%\begin{figure}[h]
%\centering
%\includegraphics[width=20pc]{Spectral-Kbar.eps}\hspace{2pc}%
%\begin{minipage}[b]{14pc}\caption{\label{fig:Kbar-spectral}Left: $\bar K$ spectral function for two different momenta at $\rho=\rho_0,2\rho_0$ and different temperatures in the range $T\leq 150$~MeV \cite{Tolos:2008di}. Right: Temperature evolution of the in-medium vector meson width associated to the modes $\bar K^*\to \bar K \pi$, $K^*\to K\pi$ at nuclear densities $\rho=\rho_0, 2\rho_0$ \cite{Ilner}.}
%\end{minipage}
%\end{figure}

\begin{figure}[h]
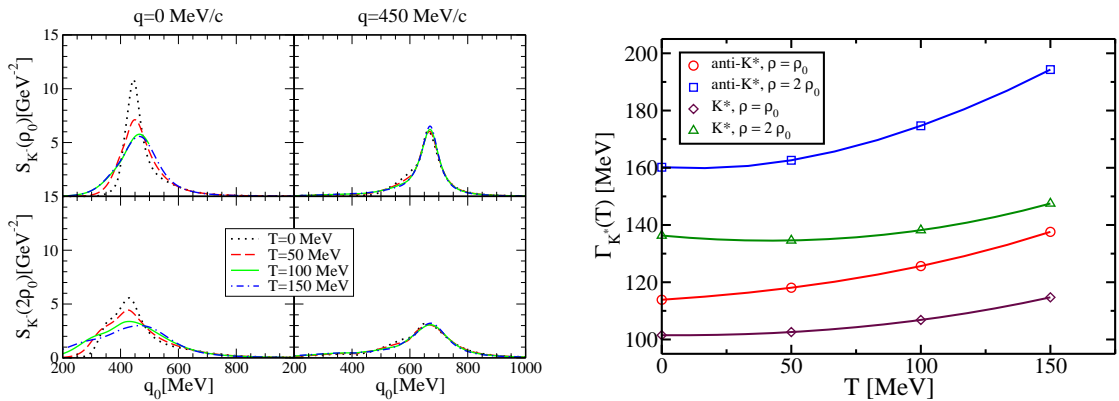

\centering
\includegraphics[width=0.44\textwidth]{spectral_tot_s+p_kbarn.v2.eps}
\hspace{1pc}
\includegraphics[width=0.43\textwidth]{gamma-Kstar-and-Kstarplus.Tdep.eps}
\caption{\label{fig:Kbar-spectral}Left: $\bar K$ spectral function in hot nuclear matter at different temperatures \cite{Tolos:2008di}. Right: Temperature evolution of the $\bar K^* (K^*)$ width due to $\bar K (K) \pi$ in-medium decay \cite{Ilner}.}
\end{figure}

The $\bar K$ spectral function, $S_{\bar K}(q_0,\vec{q};T)=-\pi^{-1}\textrm{Im}\ [q_0^2-\vec{q}\,^2-m_K^2-\Pi_{\bar K}(q_0,\vec{q};T)]^{-1}$, is shown in Fig.~\ref{fig:Kbar-spectral}~(left) at $\rho=\rho_0,2\rho_0$ and different temperatures. Far from staying as a narrow quasi-particle, the $\bar K$ exhibits a complex structure as a result of the interference of the quasi-particle peak with the $\Lambda (1405)$ and the $p$-wave $YN^{-1}$ excitations.
%The latter, being sub-threshold, partly compensate the attraction induced by the $\Lambda (1405) N^{-1}$ mode and are responsible for an appreciable low-energy tail at finite kaon momentum.
Increasing the temperature smears the Fermi surface of nucleons, reducing the attraction felt by the $\bar K$ and thus bringing the quasi-particle peak closer to the free position, whereas the more substantial mixing between the different $YN^{-1}$ modes spreads the spectral function over a wider range of energies.
The kaon ($K$), in contrast, survives as a narrow state in the nuclear medium and only at high densities and temperatures some broadening of the $K$ peak is observed.
A repulsive $K$ mass shift of about 30~MeV was obtained in cold matter at normal density, close to the result within a low-density $t \rho$ approximation.

The results discussed above constitute a microscopic input for transport simulations implementing the production of strange hadrons in HICs. Such a study has been performed, for instance, within the HSD and IQMD approaches at SIS energies \cite{Hartnack:2011cn,Cassing:2003vz}, the former one featuring full off-shell strange meson dynamics \cite{Cassing:2003vz}. 
Further work along these lines is in progress in order to account for a full catalogue of in-medium cross sections for the relevant processes involving strange hadrons. We refer to \cite{Hartnack:2011cn} for a report on strangeness production close to threshold in HICs, including predictions for recently proposed angular observables.

\subsection{Strange vector mesons in hot and dense matter}

Strange vector meson resonances ($\bar K^*, K^*$) have triggered substantial attention recently, both theoretically and experimentally. Unlike their unflavoured partners, they do not decay in the dilepton channel, making their experimental detection less clean. Still, both the RHIC low energy scan programme and the HADES experiment in GSI are currently performing measurements in order to extract the in-medium properties of hadronic resonances and, in particular, of strange vector mesons \cite{Blume:2011sb,Agakishiev:2013nta}. On the theory side, the properties of $\bar K^*$ have been studied in cold nuclear matter \cite{Tolos:2010fq} starting from a model of the $\bar K^* N$ interaction within the hidden gauge formalism (for a recent review see \cite{Oset:2012ap}).
The $\bar K^*$ selfenergy receives two contributions associated to its dominant decay mode $\bar K^* \to \bar K \pi$ and to the quasi-elastic reaction $\bar K^* N \to \bar K^* N$ and related absorption channels. The latter mechanism induces a strong broadening of the $\bar K^*$ spectral function as a result of the mixing with two $J^P=1/2^-$ states, the $\Lambda(1783)$ and $\Sigma(1830)$, which are dynamically generated in a parallel way to the $\bar K N$ interaction. At $\rho=\rho_0$, the $\bar K^*$ width is enlarged beyond 200~MeV, the in-medium $\bar K \pi$ cloud contributing about 100~MeV.
%The $\bar K^*$ nuclear optical potential reflects a moderate attraction of about 50~MeV.

The situation for the $K^*$ meson is similar to that of the $K$ and one expects milder changes in the presence of a nuclear medium. The lack of resonant states close to threshold advises to use a $t\rho$ approximation for the $K^*$ selfenergy. The calculation within the hidden gauge formalism is straightforward and reads \cite{Ilner}
\begin{eqnarray}
\label{eq:SelfKstar}
\Pi_{K^*N}=\frac{1}{2} (t_{K^* p} + t_{K^* n}) \rho_0 \frac{\rho}{\rho_0} \simeq 0.22 \, \frac{m_K}{m_{K^*}} m_{K*}^2\, \frac{\rho}{\rho_0} \ .
\end{eqnarray}
The latter equality is obtained with the lowest order tree-level amplitudes from the $K^*N$ Lagrangian evaluated at threshold, leading to a positive $K^*$ mass shift of about 50~MeV at $\rho=\rho_0$. When full unitarized scattering amplitudes are implemented, a 20\% reduction is obtained over the tree-level result, $\Delta m_{K^*} (\rho_0)\simeq 40$~MeV.

Work is being pursued to extend the former studies to a scenario of nuclear matter at finite temperature. We briefly discuss some results on the two-meson cloud contribution to the in-medium width of strange vector mesons, whereas the full calculation including also the collisional part of the selfenergy will be reported elsewhere. The in-medium width can be obtained straightforwardly from the imaginary part of the selfenergy ($\Gamma_{\bar K^*\to\bar K \pi} = - \textrm{Im}\, \Pi_{\bar K^*\bar K \pi} / P^0$), which for a $\bar K^*$ at rest reads
\begin{eqnarray}
\label{eq:ImPi}
\textrm{Im}\, \Pi_{\bar K^*\bar K \pi}(P^0,\vec{0}\,) = 
-2\pi\, g^2 \int \frac{\textrm{d}^3q\, \vec{q}\,^2}{(2\pi)^3} 
\int_0^{P^0} \textrm{d}\omega  \,
S_{\bar K}(\omega,\vec{q}\,) \,  S_{\pi}(P^0-\omega,\vec{q}\,) \, \lbrack 1+f_{\omega}+f_{P^0-\omega} \rbrack
+ \textrm{diff.}
\end{eqnarray}
The coupling $g$ is fixed to reproduce the $\bar K^*$ vacuum decay width and $f_{\omega}$ stands for the Bose distribution function (similar expressions are obtained for $K^*$ and $\phi$ mesons). The omitted terms, labelled as \emph{diff}, stand for diffusive contributions in which the $\bar K^*$ can be absorbed by a thermally excited meson, namely $\bar K^* \pi \to \bar K$ and $\bar K^* K \to \pi$.
%They are small at the nominal $\bar K^*$ mass but contribute notably at low (off-shell) energies.
%For the temperature range of interest ($T \lesssim 150$~MeV) the Bose enhancement factor plays a little role for "vacuum-like" $\pi$ and $\bar K$. However, it becomes important when these states develop low-energy modes as a consequence of their in-medium dressing.
The Bose enhancement factor becomes important when the light pseudoscalars develop low-energy modes as a consequence of their in-medium dressing.
The evolution of the $\bar K^*$ and $K^*$ in-medium decay widths with temperature is depicted in Fig.~\ref{fig:Kbar-spectral}~(right). It is worth mentioning that the HADES collaboration have recently reported a reduction of the $K^{*0}$ yield in Ar+KCl collisions at 1.76~$A\,$GeV as compared to estimations within the UrQMD transport approach \cite{Agakishiev:2013nta}.

\section{Heavy-flavour relaxation in a hadronic medium}

Given the large masses of $D$ and $B$ with respect to other relevant scales, the transport coefficients of these mesons are usually evaluated within a Fokker-Planck equation approach.
Two distinct scenarios can be addressed regarding the nature of heavy meson interactions with the hadronic ambient: $\mu_B \simeq 0$ (vanishing baryochemical potential), where the medium is predominantly composed of light mesons; and $\mu_B \ne 0$, where the heavy-meson nucleon interaction sets in and most likely drives the relaxation mechanism \cite{Tolos:2013kva}. The latter situation is expected to be encountered at low and intermediate energy HICs and in nuclear production reactions (FAIR). We focus here on the first scenario, whereas the second one has been discussed in \cite{Torres}.

%An essential ingredient to describe heavy flavour relaxation in a hadron gas is to construct a realistic model for the interaction of heavy mesons with the constituents of the medium, as constrained as possible by the available phenomenology.
An important piece of information is the fact that the spectrum of excitations of the stable open-charm mesons contains two broad resonances with dominant pionic decays in the $s$-wave: $D_0(2400)\to D\pi$ and $D_1(2427)\to D^*\pi$ (a similar situation is found in the bottom sector). This means that the cross section of the interaction of a $D^{(*)}$ meson with the most abundant constituent of the hadronic medium is dominated by a resonance at low energies. Since a resonant interaction is bound to produce shorter thermalization times, accounting for this feature is important to produce a realistic estimation of the transport coefficients.

The authors of Ref.~\cite{He:2011yi} have evaluated the drag coefficient of $D$ mesons by constructing the scattering amplitudes off $\pi$, $K$, $\eta$, $\rho$, $\omega$, $K^*$ mesons and $N$, $\Delta$ baryons in a phenomenological way, using either Breit-Wigner parameterizations or empirical data. In Refs.~\cite{Ghosh:2011bw,Abreu:2011ic,Das:2011vba,Abreu:2012et} an effective field theory constrained by chiral and heavy-quark symmetries has been used to determine the low-energy interactions of $D$'s and $B$'s  with the $SU(3)$ octet of light pseudoscalar mesons (see also \cite{Fuchs:2004fh}). An important aspect was introduced in \cite{Abreu:2011ic,Abreu:2012et} by demanding that the scattering amplitudes satisfy exact unitarity in partial waves, an essential requirement in order to extend the applicability of the low-energy theory up to temperatures $T \simeq m_{\pi}$. With a minimal set of parameters the unitarized theory dynamically generates the heavy-light meson $s$-wave resonances in good agreement with the available experimental data both in the charm and bottom sectors.

\begin{figure}[h]
\includegraphics[width=0.33\textwidth]{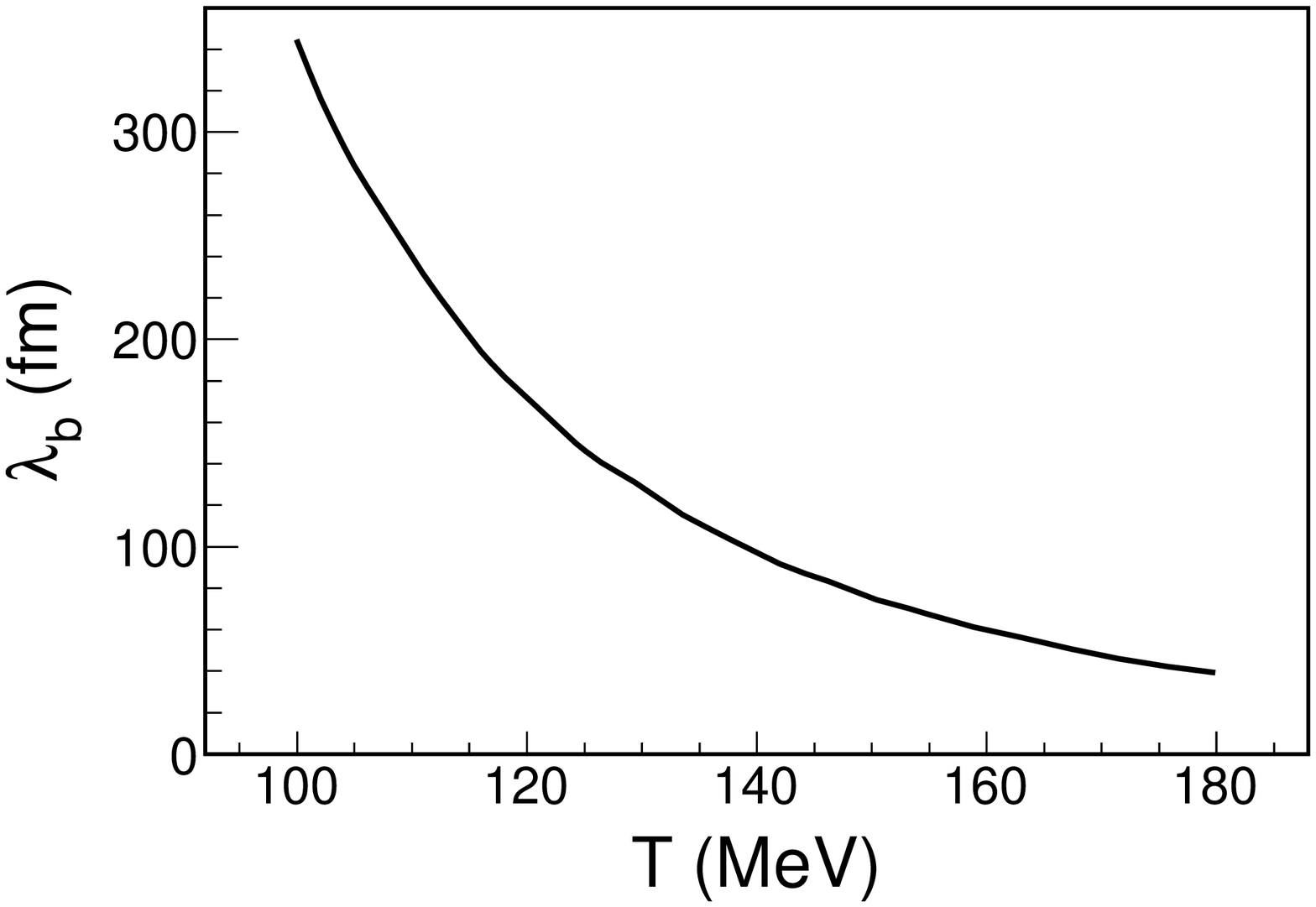}
\includegraphics[width=0.33\textwidth]{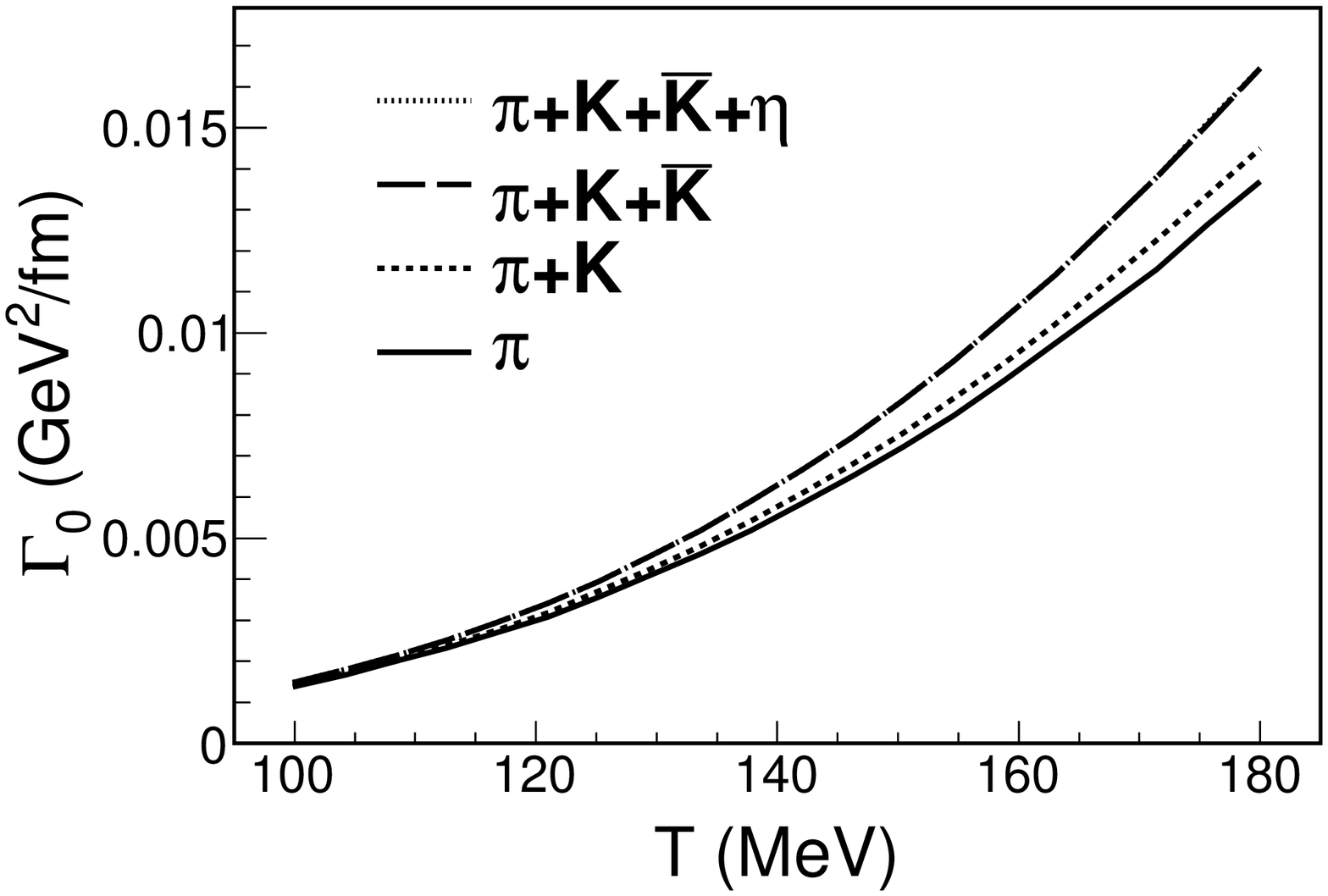}
\includegraphics[width=0.33\textwidth]{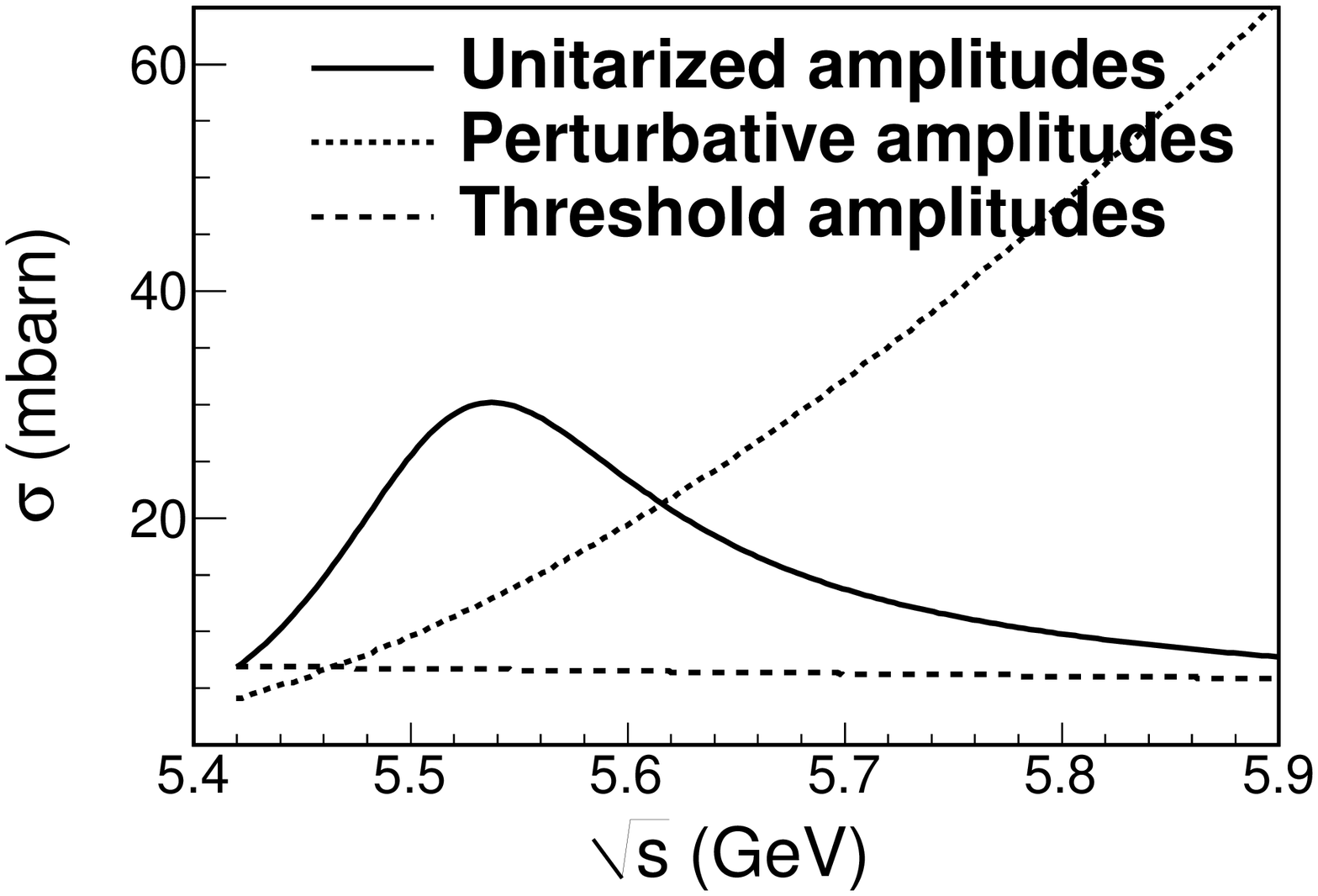}
\caption{\label{fig:coeffs}Relaxation time (left) and longitudinal diffusion coefficient (middle) of bottomed mesons at $p=0.1$~GeV$/c$ as a function of temperature \cite{Abreu:2012et}. The total elastic $B\pi$ cross section is shown in the right panel (the solid curve exhibits a $B_0$ resonance at $\sqrt{s}\simeq 5.53$~GeV).}
\end{figure}

The (inverse of the) drag coefficient $F$ and the longitudinal diffusion coefficient $\Gamma_0$ at $p=100$~MeV$/c$ for bottomed mesons are shown in Fig.~\ref{fig:coeffs} (left and middle panels, respectively), as obtained in \cite{Abreu:2012et} . Both drag and diffusion are considerably strengthened in the hotter stages of a HIC, with significant interactions between heavy mesons and the thermal medium (for results in the charm sector see \cite{Torres}). As expected, the most relevant contribution comes from the pion gas, followed by kaons and (mostly) anti-kaons. At $\mu_B = 0$, the effect of baryons (nucleons and $\Delta$'s) is negligible even at the highest considered temperatures. The relaxation time of bottomed mesons travelling with 1~GeV$/c$ momentum is found $\lambda_B\simeq 80$~fm at $T=150$~MeV, still much greater than the typical lifetime of the hadron gas, thus rendering bottomed mesons as optimal carriers of information of the phase transition upon exiting the interacting region. Comparatively, charmed mesons are more affected by the interactions with the medium ($\lambda_D\simeq 40$~fm at the same conditions), particularly at finite baryochemical potential \cite{Tolos:2013kva}.
%While not completely relaxing during the lifetime of the collision, they may partly loose memory of the initial stages.
The role of unitarization within the heavy-meson chiral effective approach is illustrated in the right panel of Fig.~\ref{fig:coeffs}.

%Finally, the right panel of Fig.~\ref{fig:coeffs} exhibits the 

\vspace{-0.6cm}
\section{Summary}

We have presented a selection of results and discussed work in progress concerning the properties of strange and heavy mesons in hot and dense strongly interacting matter. We have emphasised the benefits of unitarized coupled-channel approaches based on the chiral Lagrangian, which provide the essential features of the interaction of these mesons with the hadronic constituents of the medium in a highly predictive way. The use of realistic hadronic interactions accounting for medium effects in conjunction with microscopic transport approaches qualifies as a powerful device to investigate the complicated reactions that take place in heavy-ion collisions.

\vspace{-0.2cm}
\section*{Acknowledgments}
DC thanks the organizers of FAIRNESS 2013 for a stimulating conference. Work supported by grants BMBF (Germany) no.~05P12RFFCQ and MINECO (Spain) no.~FPA-2011-27853-C02-02.

\end{document}